\documentclass[a4,11pt]{llncs}
\usepackage[left=4cm,right=4cm,top=3cm,bottom=3cm]{geometry}
\usepackage{enumerate}
\usepackage{tabularx}
\usepackage{graphicx}
\usepackage{listings}
\usepackage{algorithm}
\usepackage{algorithmic}
\usepackage{amsfonts}
\usepackage{color}
\usepackage{caption}
\usepackage{subcaption}
\usepackage{multirow}
\definecolor{DBlue}{rgb}{0,0,0.4}
\definecolor{DRed}{rgb}{0.4,0,0}
\definecolor{CGreen}{rgb}{0.3,0.7,0.4}
\definecolor{OliveGreen}{rgb}{0.3,0.7,0.4}
\definecolor{LGrey}{rgb}{0.95,0.95,0.95}

\usepackage{url}
\usepackage{hyperref}

\usepackage[]{listings}

\lstdefinelanguage{scala}{
       morekeywords={
                override, try, catch, throw, private, public, protected, import, package, implicit, final, package, trait, type, class, val, def, var, if, this, else, extends, with, while, new, abstract, object, case, match, sealed, for, yield,super},%
         sensitive=t, %
   morecomment=[s]{/*}{*/},morecomment=[l]{\//},%
   escapeinside={/*\%}{*/},%
   rangeprefix= /*< ,rangesuffix= >*/,%
   tabsize=2,%
   morestring=[d]{"}%
 }

\begin{document}

\title{Kevoree Modeling Framework (KMF):\\ Efficient modeling techniques for runtime use}

\titlerunning{Kevoree Modeling Framework (KMF)}  

%
\author{Fran\c{c}ois Fouquet$^1$ \and Gr\'egory Nain$^1$ \and Brice Morin$^2$ \and Erwan Daubert$^3$ \and Olivier Barais$^3$ \and
No\"{e}l Plouzeau$^3$  \and Jean-Marc J\'ez\'equel$^3$}
\authorrunning{F. Fouquet and al.}   
%
%
\institute{University of Luxembourg \newline
Interdisciplinary Center for Security, Reliability and Trust (SnT)\newline
Luxembourg, Luxembourg \newline
\{Firstname.Name\}@uni.lu
\and
SINTEF \newline
Oslo, Norway \newline
\{Firstname.Name\}@sintef.no
\and
University of Rennes 1, IRISA, INRIA Centre Rennes \newline
Campus de Beaulieu, 35042 Rennes, France \newline
\{Firstname.Name\}@inria.fr
}

\maketitle              

\begin{abstract}
\vspace{-15em}
The creation of Domain Specific Languages (DSL) counts as one of the main goals in the field of Model-Driven Software Engineering (MDSE).
The main purpose of these DSLs is to facilitate the manipulation of domain specific concepts, by providing developers with specific tools for their domain of expertise.
A natural approach to create DSLs is to reuse existing modeling standards and tools.
In this area, the Eclipse Modeling Framework (EMF) has rapidly become the \textit{defacto} standard in the MDSE for building Domain Specific Languages (DSL) and tools based on generative techniques.
However, the use of EMF generated tools in domains like Internet of Things (IoT), Cloud Computing or Models@Runtime reaches several limitations.
In this paper, we identify several properties the generated tools must comply with to be usable in other domains than desktop-based software systems. 
We then challenge EMF on these properties and describe our approach to overcome the limitations.
Our approach, implemented in the Kevoree Modeling Framework (KMF), is finally evaluated according to the identified properties and compared to EMF.

\keywords{Domain Specific Language, Model Driven Software Engineering,\\ Kevoree Modeling Framework, Models@Runtime, EMF, adaptation}

~\\
Technical report, University of Luxembourg, May 2014\\
ISBN 978-2-87971-131-7/TR-SnT-2014-11 
\end{abstract}

\pagebreak


\section{Introduction}
\label{sec:intro}

Model-Driven Software Engineering investigates how models, as abstract representations, can support and enhance the authoring of software systems.
In particular, Domain-Specific Modeling Languages (DSMLs) aim at providing software developers with dedicated languages and tools specifically created for their business domain.
DSMLs usually leverage an Object-Oriented software API, generated from a metamodel, facilitating the authoring and exchange of domain-specific concepts between several tools, applications and stakeholders.
This API can for example support graphical or textual editors to edit models, or specific load and save algorithms to serialize and exchange models.
The primary focus of DSMLs was to ease the comprehension and manipulation of concepts during the design phase of an application. 
In this area, the Eclipse Modeling Framework (EMF) has rapidly become the \textit{defacto} standard in the MDSE for building DSMLs and associated tools using generative techniques.
As the boundaries between design-time and runtime become more and more blurry, DSMLs are increasingly embedded at runtime to monitor, manage, maintain and evolve running applications, leading to the emerging Models@Runtime paradigm~\cite{BlairBF09}.

However, the integration and use of EMF APIs and tools in a Models@Runtime context reached several limitations.
For example, when applied to the domain of the Internet of Things~\cite{thingml}, the execution environments are usually constrained in terms of memory and computational power, making it difficult to embed the DSML tools generated by EMF.
When applied to the Cloud Computing domain, the models created with a specific DSML for cloud are extensively exchanged, cloned and compared in order to manage the maintenance and adaptation of large-scale distributed applications. These operations call for very efficient means to clone and access model elements, which EMF fails at providing.
Also, the dynamic adaptation of a software system ({\em e.g.} to manage the scalability of Cloud-based system), requires efficient search and validation techniques to maximize the continuity of service, and minimize the downtime when reconfiguring the system.

This paper originates from the application track of MODELS'12~\cite{Fouquet:2012fk}. Our main contribution is to integrate best development practices, largely accepted in the Software Engineering community and largely disregarded in the MDSE community, in order to get improved performances on the generated code, compared to standard modeling tools and frameworks (and EMF in particular). This contribution is integrated into the Kevoree Modeling Framework (KMF), an alternative to EMF still compatible.

This paper extends our previous work in three main aspects:
\begin{enumerate}[i)]
	\item It elicits the requirements and the state of practice in using DSLs and models during the execution of a software, in three domains.
	\item It highlights the last improvements made in KMF to improve efficiency. In particular it focuses on the new features introduced to decrease the memory footprint, to decrease the time required to load or save models, to support a partial clone operator and to query a find model elements in a model using ``KMF Query Language - Path Selector''.
	\item It evaluates KMF and EMF against the properties identified as required to use models during runtime in several domains.
\end{enumerate}

The outline of this paper is the following.
Section~\ref{sec:requirements} presents use cases in three different domains, in order to illustrate the requirements raised by the use of models as first-class entities at run-time.
Section~\ref{sec:sota} gives a state of practice of using design-time tools like EMF for runtime manipulations of models.
The contribution of this paper, the Kevoree Modeling Framework(KMF), is described in Section~\ref{sec:contrib}.
Section~\ref{sec:evaluation} evaluates our alternative implementation in comparison to EMF.
Section~\ref{sec:conclusion} concludes this work and presents future work.


\section{Requirements for using models at runtime}
\label{sec:requirements}

This section introduces briefly three domains in which models are used during the system runtime. For each domain, a specific DSL is used and highlights different requirements for the object-oriented API and the associated tools generated. Then, several properties are identified and listed as mandatory to enable a consistent use of models during runtime.

\vspace{-1em}
\subsection{Use cases}
\label{sec:usecases}
\vspace{-1em}
We present in this section three cases in which the use of models during runtime, for different purposes, requires some properties to be guaranteed by the tools generated for the DSL.

\subsubsection{Internet of Things}~\\
The domain of the Internet of Things relates to the interconnection of communicating object. 
The variability and heterogeneity of devices to be considered in this domain challenges the possibilities of interoperate the devices and the development of software systems to offer new services.
In this domain, models are used during the software execution to reflect the configuration of devices, their links to siblings or services. 
Also, the sporadic presence of devices requires dynamic adaptations of the running system to install and remove pieces of software to control each specific device present in the range of control. 
The models are very useful in this context to describe what are the services available on each device, where are the binaries to connect it, etc.
In addition, model structures are useful to express composition of services operations, inherently necessary to compose data collected asynchronously.

The main concern here is about the memory restrictions and time of adaptations. Indeed, the kind of execution environment used in this domain are usually constrained in terms of memory and computational power. 
Thus, attention has to be paid to the size of the generated code and the memory necessary for its execution, as long as to the time necessary to read (load) models.

\subsubsection{Cloud Computing}~\\
Cloud computing is characterized by a dynamic provisioning of services on computation resources.
Also called elasticity, this intelligent process must dynamically place several software to adapt the computation power to the real consumption of customers' software.
However, costs limits imposed by cloud customers force the elasticity mechanism of cloud to perform some choices, sometime contradictory.
For instance, cost and power can not be optimize at same time, because of their intrinsic dependency.
This leads to multi-objective optimisations, which consider lot of information to compute a placement of software offering the best tradeoff.

In this context, models are perfect candidates to represent the state of the cloud infrastructure and ease the work of these optimisation algorithms.
Starting from the model reflecting the current state of the cloud infrastructure, these algorithms explore and evolve lot of configurations to find the most suitable tradeoffs between all cloud metrics.
In this process, the high number of models (configurations as used by Ferry~\textit{and al}~\cite{ferry2013towards}) created and dropped calls for very efficient mechanisms for allocation, deletion, clone and mutation, because they constitute the basic operations of evolutionary algorithms~\cite{frey2013search}.

\subsubsection{Software systems self-adaptation using Models@Runtime}~\\
Models@runtime is an emerging paradigm aiming at making the evolution process of applications more agile and dynamic, still leveraging the key benefits of modeling: simplicity, efficiency and safety.
It is inspired by solid research in reflective software engineering~\cite{DBLP:conf/reflection/2001} ({\em e.g.} Meta-Object Protocol~\cite{Kiczales1991}) and makes available, at runtime, an abstraction of the system for the system itself or for external human/programmatic actors. This abstraction helps in supporting efficient decisions related to the adaptation or evolution of the system by hiding (simplifying) irrelevant details.

Kevoree~\cite{DBLP:conf/models/MorinFBJSDB08,Fouquet:2012:DCM:2304736.2304759} is an open-source project leveraging the Models@Runtime (M@RT) paradigm to support the continuous design of complex, distributed, heterogeneous and adaptive systems. The overhead implied by this advanced reflection layer (Models@Runtime) has to be minimized to reduce as much as possible the resources (computation and memory) mobilized, and should enable the efficient dissemination of models among nodes~\cite{fouquet:hal-00688707}, to manage the widest possible types of node (including Android, Java Embedded).
If the M@RT overhead is too heavy, the process tends to be centralized, impacting the overall resilience of the software system (single point of failure, etc).\\

Current MDE tools were however designed and engineered without any consideration for runtime constraints, since they were originally thought for design-time use only, where time and resource constraints are much softer. The following section gives a more detailed description of the properties to comply with when targeting a use at runtime.

\subsection{Generic requirements for using models at runtime}
\label{sec:generic:requirements}

In the three domains presented in~\ref{sec:usecases}, applications need to perform common operations on models, where performance and safety are critical issues.
The memory footprint is the first performance concern. 
Indeed, the generated Object-Oriented API and all its dependencies must be compatible with the hosting device in terms of memory footprint, to enable the inclusion on low-resource IoT nodes.
Secondly, models should come with efficient (un)-marshaling techniques to enable the rapid exchange of models between distributed cloud nodes.
Also, model cloning and concurrent, yet safe, manipulations of models should be supported by the framework to enable distributed reasoning, {\em e.g.} based on exploratory techniques such as genetic algorithms for cloud management.
Finally, models should be traversable efficiently to enable rapid computation on already known elements, for comparison or merge of models for instance, which are basic operations.
Each of these key requirements for M@RT is further described in the following sections.

\subsubsection{Reduced memory footprint}
~\\
The memory overhead introduced by models at runtime has to be considered from both static and dynamic point of view.\\

The \textbf{static memory footprint} basically depends on the complexity (size) of the metamodel that defines the concepts and relationships among these concepts to be manipulated in the models.
Basically, the more complex a metamodel is, the bigger the generated Object-Oriented(OO) API is.
Also, metamodels themselves conform to a meta-metamodel (M3), such as Executable Meta-Object Facility (EMOF)\cite{OMG:2013} or ECORE, which can also participate in the static memory footprint. However, the footprint of the M3 remains stable independently of the size of the metamodels.\\
The number and size of the dependencies required by the generated OO API have to be minimized, because each device must provision these dependencies to be able to run the API.
Heavy dependencies would indeed increase the initialization or update time of a device and reduce the memory space available for the models.

The \textbf{dynamic memory footprint} required to manipulate a model using the generated OO API is linked to the size of the manipulated model.
Thus, the number and size of the objects required to instantiate a model element ({\em i.e.} the concept and its relationship) have to be watched, because of their impact on the number of model elements that can be kept in memory.
If this footprint is too high for a device, the model will be swapped ({\em i.e.} unloaded from memory to disk), consuming a lot of resources to perform this operation.
All this implies that the memory used to create a model element has to be controlled to allow the manipulation of big models on devices with limited resources.

\subsubsection{Efficient models (un)marshalling}
\begin{figure}[h!]
\centering
\includegraphics[scale=0.40]{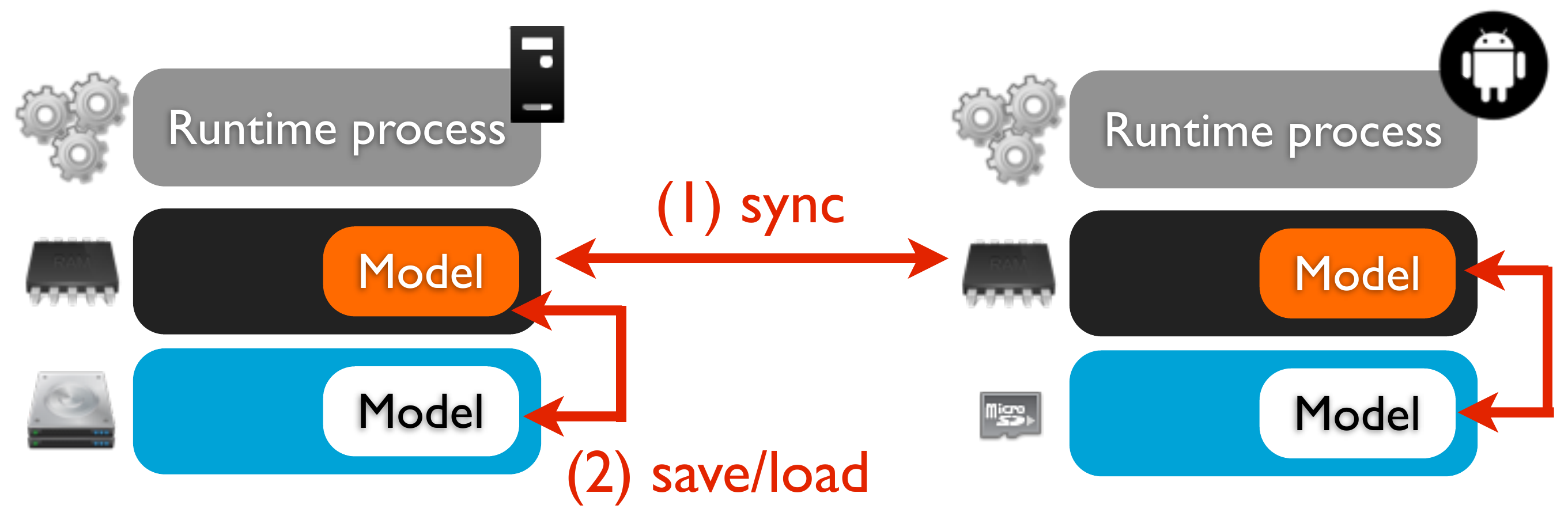}
\caption{Model@Runtime (un)marshalling operation case study illustration}
\label{fig:MARSaveLoad}
\vspace{-2.5em}
\end{figure}

~\\
The efficiency of marshaling and unmarshaling operations is a key criteria for models@runtime, because they are the basic input/ouput operations on a model at runtime, and should not ``freeze'' the system while they are serialized or loaded.
Models are used in memory for validation, verification and system manipulation. 
They are also extensively exchanged across local processes and remote nodes to disseminate decisions, and stored to disk for later use ({\em e.g.} system restore, {\em post mortem} analysis as with a flight data recorder (black box)).
To describe this requirement, the Figure~\ref{fig:MARSaveLoad} depicts two critical use cases of models I/Os ({\em i.e.} (un)marshalling operations) in Kevoree. The first is the exchange of models across nodes (1) to build a shared reflexive layer; the second concerns backup and restore operations of models (2).

\vspace{-2em}

\subsubsection{Concurrent read/write usage of models}
~\\
When used at runtime, models are generally at the centre of a highly concurrent environment.
In the IoT domain for instance, different probes integrated in several devices can update a shared context model. Obviously, this shared model must offer safe and consistent read and write operations.
\vspace{-1em}
\begin{figure}[h!]
\centering
\includegraphics[scale=0.37]{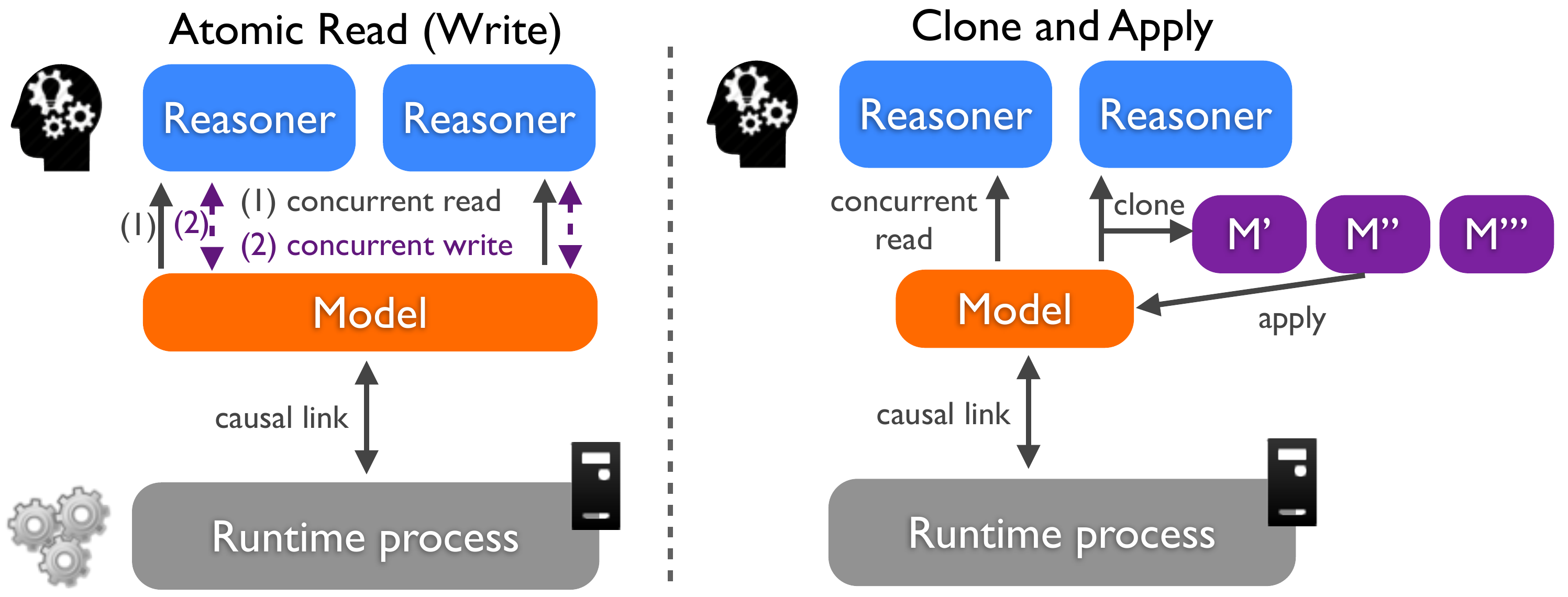}
\caption{Shared Model@Runtime protection strategies}
\label{fig:MARSharedOps}
\vspace{-1em}
\end{figure}
Like shared memory~\cite{chase1994sharing}, the generated OO API must implement a protection strategy in order to offer safe and efficient concurrent read or concurrent read-and-write operations for the various producers writing in the model and consumers of these data.

In case of a concurrent read strategy, the generated API must ensure that the multiple threads can reliably, independently and consistently access and iterate over model elements.
Similarly in case of a concurrent write strategy, the framework must ensure the atomicity of each modification.
This is depicted on the left side (Atomic Read/Write) of Figure~\ref{fig:MARSharedOps} where {\it Reasoners} components are both producers and consumers.

The cloning of models is at the centre of this concurrency problem. 
It acts as a concurrent protection for each producer/consumer, which works on its own clone of the current model, as presented in Figure~\ref{fig:MARSharedOps} that depicts a Cloud management use case where search-based algorithms use small mutations as basic operations to explore a space of potential configurations.
The memory and computational efficiency of the cloning strategies are thus key requirements for the generated API.

\subsubsection{Efficient navigation across model elements}
\label{sec:generic:nav}

\begin{figure}[hb!]
\centering
\begin{subfigure}[b]{.6\textwidth}
\centering
\includegraphics[width=\textwidth]{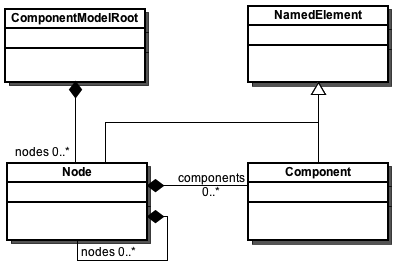}
\caption{Kevoree Metamodel exerpt}\label{fig:kevoree_MM_path}
\end{subfigure}%
\begin{subfigure}[b]{.4\textwidth}
\centering
\includegraphics[width=.8\textwidth]{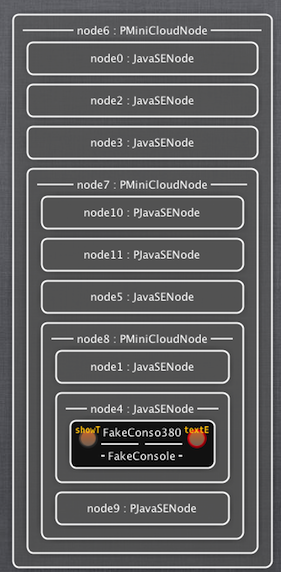}
\caption{Kevoree Model example}\label{fig:kevoree_Model_path}
\end{subfigure}
\caption{Illustration of the case with Kevoree}\label{fig:kevoree_pathexample}
\end{figure}

The purpose of a DSL is to offer a \textbf{simple} way to represent and manipulate domain specific models.
Models can create very complex graph structures in which the navigation is complicated. 
So complicated, that iterating over relations and model elements in order to find a specific element create serious issues on performance and code complexity.
To exemplify this problem, we consider a small excerpt of the Kevoree meta-model (Fig.~\ref{fig:kevoree_MM_path}) which is used to organize the placement of components and hosted nodes, on execution nodes.
With this meta-model one can easily build the Kevoree model presented in figure~\ref{fig:kevoree_Model_path}, where a component instance is hosted on a node called 'node4' itself hosted on 'node8', hosted on 'node7', hosted on 'node6'.\\
Let now imagine the software system is trying to place this component instance on a node according to several constraints. 
To perform this task, several threads are running different algorithms on clones of the model. 
The need for an efficient navigation becomes obvious when the system has to compare, select and merge the solutions from all threads.

\vspace{-1em}
\section{State of the practice: Using EMF to generated DSLs for runtime usage}
\label{sec:sota}

A natural way to create a DSL is to rely on tools and techniques well established in the MDE community, and in particular, the {\em de facto} EMF standard.
This section provides a brief overview of EMF and then discusses the suitability of this modeling framework to generate a Object-Oriented API usable at runtime, with respect to the requirements identified in the previous section.

\vspace{-1em}
\subsection{Overview}
\par EMF is a Java-based EMOF implementation and code generation facility for building languages, tools and other applications based on a structured data model.
From a metamodel conforming to the Ecore meta-metamodel (which is largely inspired by/aligned with the EMOF meta-metamodel), EMF provides tools and support to create Domain Specific Languages (DSLs) on top of the Eclipse IDE. Additional languages and tools like EMFText or GMF make it possible to rapidly define textual or graphical syntaxes for these DSLs.\\

\begin{figure}[h!]
\vspace{-2em}
\center
\includegraphics[width=0.8\textwidth]{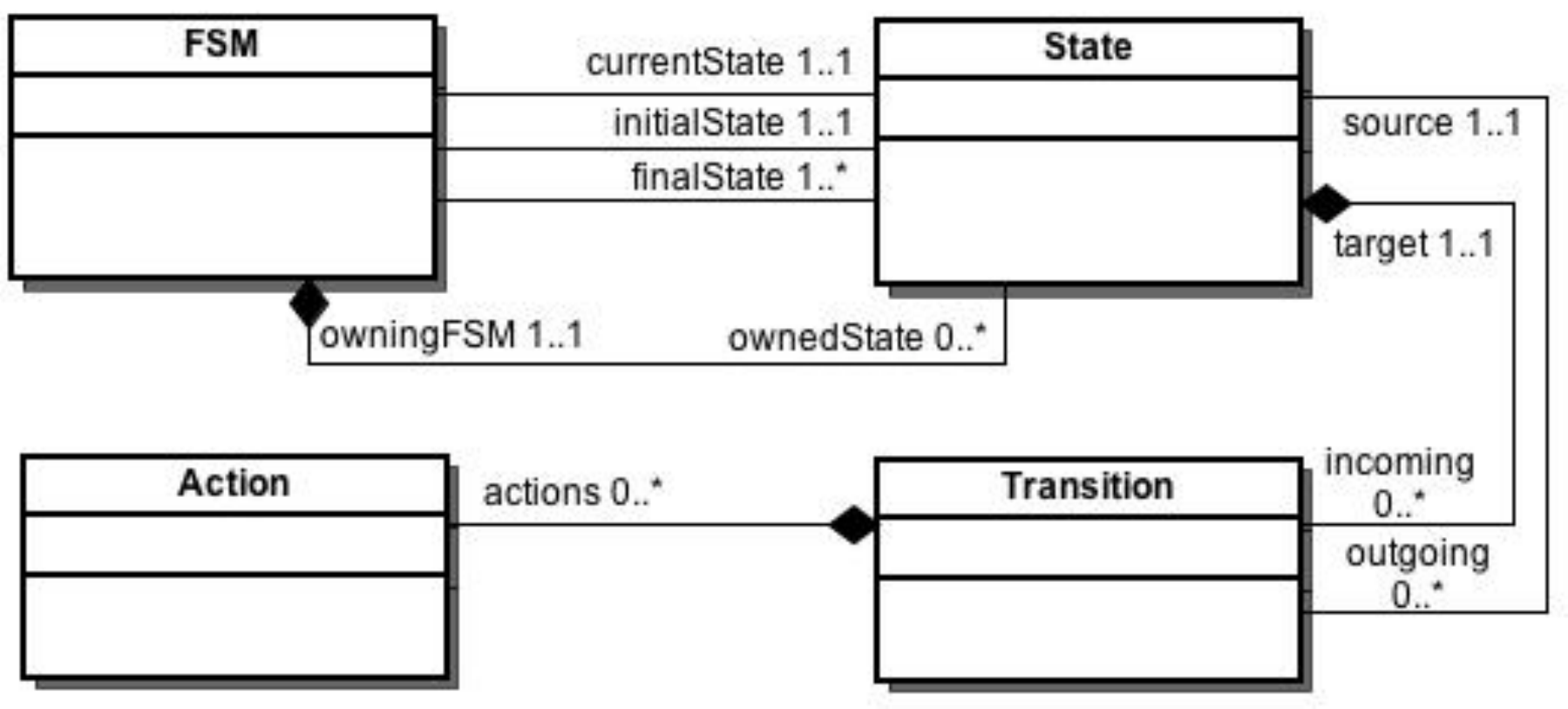}
\caption{Finite State Machine Metamodel used for Experiments}\label{fsmmeta}
\vspace{-2em}
\end{figure}

In order to evaluate the code generated by EMF, we consider one of the simplest well known meta-model within the MDE community: the Finite State Machine (FSM).
This metamodel (Fig.~\ref{fsmmeta}) is composed of four meta-classes: FSM, State, Transition and Action.
A FSM contains the States of the machine and references to initial, current and final states.
Each State contains its outgoing Transitions, and Transitions can contain Actions.

To measure the dynamic memory usage and time consumption on the different requirements, we use the code generated  by EMF to programmatically create a flat FSM (each state has exactly one outgoing transition and one incoming transition, except the initial and final states) model composed of 100,000~State instances, 99,999 transitions and an action for each transition. The experimentations have been performed on a Dell Precision E6400 with a 2.5~GHz iCore~I7 and 16~GB of memory.

The following sections discuss the position of EMF {\em w.r.t.} the identified requirements.

\subsection{Memory aspects}
For this evaluation, we first generate the modeling API (source code) of the FSM metamodel with the {\it EMF genmodel} tool, in a project called \textit{testemfdependencies}.

\begin{figure}[h!]
\vspace{-2em}
\includegraphics[width=1.\textwidth]{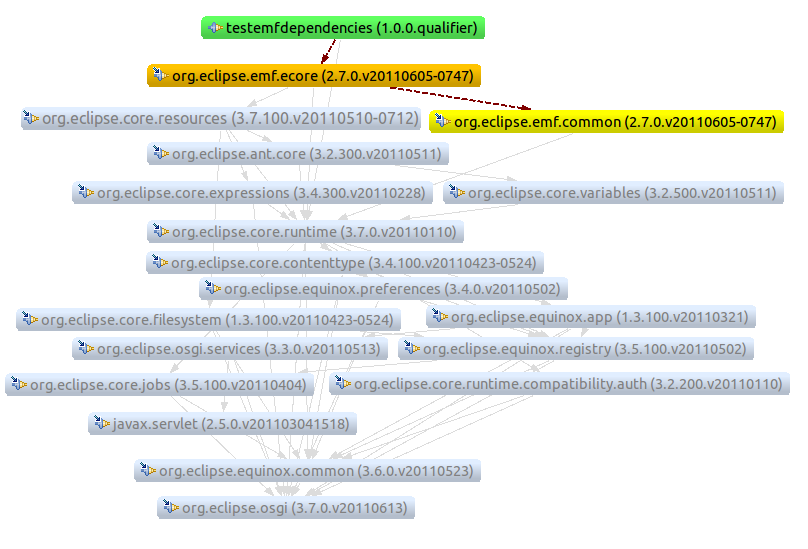}
\caption{Dependencies for each new metamodel generated code}\label{dependencies}
\vspace{-1em}
\end{figure}

Figure~\ref{dependencies} shows the dependencies needed for the generated modeling API to compile and run.
The analyse of these dependencies shows that the generated code is tightly coupled to the Eclipse environment and to the Equinox runtime (Equinox is the version of OSGi by the Eclipse foundation).
Although this is not problematic when the generated API is used in the Eclipse environment,
these dependencies are more difficult to resolve and provision in a standalone context outside Eclipse.

In the case of the FSM, the business code generated from the FSM metamodel is only {\bf 55~KB}, but requires {\bf 15~MB} of dependencies when packed in a standalone executable JAR. This represents an overhead of {\bf 99,6\%}. Moreover, the reflexive calls extensively used in the generated API makes it difficult to reduce the size with tools like ProGuard\footnote{ProGuard is a code shrinker (among other features not relevant for this paper) for Java bytecode:  http://proguard.sourceforge.net/}. Beyond the technical for memory optimization at runtime, we can extract the following generality: by introducing dynamic reflexive call (e.g dynamic because came from dynamically created string) we forbid to use dead code analyser and shrinker, then we introduce a runtime overhead. 

As for the dynamic memory, the creation of the experimental FSM model lasts for 376~ms and uses 104~MB of the heap memory.

This overhead is one of the main limitations of EMF when the generated API must be embedded at runtime.

\subsection{Load and Save operations}

EMF uses the XMI format\cite{OMG:2011fk} as default serialization strategy to allow any tool supporting this interchange format ({\em e.g.} Xtext~\cite{Merkle:2010:TMT:1869542.1869564,Eysholdt:2010:XIY:1869542.1869625}, EMFText~\cite{emftext}, GMF~\cite{gmf} or ObeoDesigner~\footnote{http://www.obeodesigner.com/}) to load/save the models.\\

EMF provides a generic loader that uses reflexive mechanisms.
It navigates the metamodel and makes reflexive calls to classes and methods of the generated API to load a model.
This loader is not generated, which limits the size of the generated code, but its genericness and reflexive approach have several drawbacks. 
First, both the model and the metamodel have to be parsed and loaded in memory, which is consuming a lot of time and memory. The reflexive calls are also very costly.\footnote{See the Java tutorial on reflection, in particular the discussion on the performance overhead: http://docs.oracle.com/javase/tutorial/reflect/}
To hide and avoid some of this algorithmic complexity, the EMF loader implements a {\em lazy loading} mechanism which actually loads the attributes and references of model elements only if the model element is actually accessed.

As for the marshaling (save) operation, the EMF standard serializer works in two steps. It first transforms the model to be serialized into a Document Object Model (DOM). The size of this temporary DOM structure is directly linked to the size of the model. Then it prints this DOM in a file.

The serialization of our experimental FSM model to a file lasts 7021~ms and the loading of the model from this file lasts 5868~ms, which is a lot considering the power of the computer used for the experimentation.

\subsection{Concurrency in the generated code}

Many runtime platforms used to support dynamic software architectures implement their own class loaders to keep control of resources. It is the case in OSGi, Frascati or Kevoree for instance.
Then, software running on these platforms must use these class loaders in order for the platform to offer an efficient and reliable dynamic class loading.
However, the use of static registries in EMF leads to incompatibilities with runtime platforms using multi-class loaders.

Moreover, EMF provides no guarantee of thread safety of the generated code~\footnote{http://wiki.eclipse.org/EMF/FAQ\#Is\_EMF\_thread-safe.3F}.
Even worse, when a collection on a model element is accessed by two threads in parallel, the iterator on the collection may be shared.

Nevertheless, this need for a proper concurrency management is essential, because several IoT sensors may access the model at the same time or several cloud-management threads may put their results in the model concurrently.
EMF is thus not adapted for this kind of use.

As for the clone of models, we used the \texttt{EcoreUtil} to clone the FSM experimental model, and the process took 3588~ms, which is not optimal.

\vspace{-1em}
\subsection{Navigation in models}

The code generated by EMF provides an embedded visitor pattern and an observer pattern~\cite{gof94}.
Nevertheless, the navigation through model elements and the lookup of a specific known element can be very costly.
To illustrate that, we consider the previously defined Kevoree component lookup example detailed in section~\ref{sec:generic:nav} (Figure~\ref{fig:kevoree_Model_path}), and use the plain Java API generated by EMF.
The code required to access the component instance element in the model is presented in Listing~\ref{lst:plainJava}.

Let consider $n$ the number of nodes at the root of the model, $o$ the number of nodes hosted by 'node6', $p$ the number of nodes in 'node7', $q$ the number of 'node8', $r$ the number of component instance in 'node4'; the algorithmic complexity of that piece of code is O($n \times o \times p \times q \times r$) which creates a combinatorial explosion with the growth of nodes at each level, and the depth of nested nodes.

\begin{scriptsize}
\begin{lstlisting}[frame=L, language=java, caption={Collect of the FakeConsole component instance in plain Java.}, label={lst:plainJava}]
for (ContainerNode node : model.getNodes()) {
  if (node.getName().equals("node6")) {
    for (ContainerNode node2 : node.getHosts()) {
      if (node2.getName().equals("node7")) {
        for (ContainerNode node3 : node2.getHosts()) {
          if (node3.getName().equals("node8")) {
            for (ContainerNode node4 : node3.getHosts()) {
              if (node4.getName().equals("node4")) {
                for (ComponentInstance i : node4.getComponents()) {
                  if (i.getName().equals("FakeConso380")) {
                    fConsole = i;break;
                  }
     [...]
}
\end{lstlisting}
\end{scriptsize}

Functional languages ({\em e.g.} Scala), or languages dedicated to model query/transformations ({\em e.g.} OCL, QVT) can of course be used instead of a plain Java approach to facilitate the writing of this code, but come with additional integration costs and worse performances, typically due to the use of an interpreter.

A common workaround for this performance issue is the creation of cache or similar buffer approaches.
Such techniques introduce new problems like eventual consistency, which are often in opposition with multi-thread access.
As a consequence, common modeling projects use helpers with high computational complexity to index and search model elements, which has serious performance penalty at runtime.

\vspace{-1em}
\subsection{Synthesis}

This section shows that MDE tools, initially developed for design-time activities can be used to create proof-of-concept Models@Runtime platforms.
However, several limitations (in terms of memory and time efficiency) and drawbacks (thread safety not guaranteed) make their use sub-optimal or even impossible at runtime.

In order to improve MDE tools for a runtime usage, and make them generally more efficient, we focus the work described in this paper on four questions:
\begin{enumerate}
	\item How to reduce the number of dependencies and intermediate objects created, to reduce the memory requirements ?
	\item How to take advantage of the information available in the metamodel to generate more efficient loaders and serializers?
	\item How to offer an efficient support for concurrent access to models at runtime?
	\item How to efficiently access a specific model element?
\end{enumerate}

In the next section, we describe the Kevoree Modeling Framework which implements our contributions to address these questions.


\section{The Kevoree Modeling Framework}
\label{sec:contrib}

This section presents the Kevoree Modeling Framework (KMF), an alternative to the Eclipse Modeling Framework to address the requirements on the generated code listed in section~\ref{sec:requirements}, imposed by the use of models at runtime .
This section is divided into four subsections according to the four key requirements.

As an introduction to the contribution, the section~\ref{sec:contrib:overview} gives an overview of the main principles that drove our work.
Section~\ref{sec:contrib:memoryFootprint} then describes how we reduced the memory footprint of the generated code. 
Section~\ref{sec:contrib:saveLoad} presents the improvements made on I/O operations (load and save) for models. 
The mechanisms to manage concurrency are explained in Section~\ref{sec:contrib:clone}. 
Finally, Section~\ref{sec:contrib:path} introduces the KMF Query Language (KMFQL) to efficiently reach a specific known element in a model.

\subsection{General approach}
\label{sec:contrib:overview}
We followed several principles and guidelines during the development of KMF in order to meet the performance requirements. Here is the list.

\begin{enumerate}[A)]
	\item {\em Avoid the creation of temporary objects as much as possible}. 
While the creation of objects in the Java virtual machine is not very costly, their destruction by the garbage collector\footnote{responsible for freeing the memory of not referenced objects} is a more complex and costly operation\footnote{See http://www.oracle.com/technetwork/java/gc-tuning-5-138395.html}.
We optimize the memory by reusing objects already loaded instead of creating temporary objects.

	\item {\em Sharing immutable objects}. In Java, several collections\footnote{See http://docs.oracle.com/javase/tutorial/essential/concurrency/immutable.html} and primitive objects can define an immutable state. 
Leveraging immutability, we automatically apply a flyweight pattern~\cite{gof94} to share these immutable objects among several models to reduce the overall memory footprint.

	\item {\em Flat reflexivity}. By relying on a sound generative approach, any modification of the metamodel implies a regeneration of all the tools, but it also allows using the closed world assumption~\cite{reiter1978closed}. Thus, model manipulations at runtime rely on a finite and well-defined set of types. This allows generating a flat reflexive layer, composed of a finite set of matching functions, avoiding costly reflexive calls at runtime.
	
	\item {\em No fancy dependency}. To be embeddable in the wider range of hosting nodes, the modeling framework must limit its dependencies. The code generated with KMF depends only on no third-party library.

	\item {\em Direct Memory Access for models I/O}. KMF uses as much as possible Direct Memory Access to load and save models from the network or the disk to the local memory. 
This principle reduces the need for buffering during model I/O operations.\\
\end{enumerate}

\vspace{-2em}
\subsection{Reduction of the memory footprint}
\label{sec:contrib:memoryFootprint}

The static footprint reduces the memory available for the dynamic creation of objects necessary for business processing. It is thus important to watch both static and dynamic memory requirements of the generated code.

\paragraph{Static memory footprint}~\\
To limit the dependencies and thus the static memory footprint, we decided to restrict the inheritance relationships of our generated code to sibling classes and elements from standard libraries only.

In a first attempt~\cite{Fouquet:2012fk}, the source code generated by KMF already allowed to save some 8~MB of dependencies, and up to 13~MB with the help of a code shrinker (ProGuard).
Now, KMF generates Kotlin code, with an even lighter static footprint.
On the same FSM metamodel, KMF generates 76~KB of code (including the core API, as well as model loader/serializer, and a query API) to be compared to the 55~KB of EMF pure code.
Including the dependencies ({\em i.e.} the Kotlin standard library), the standalone generated code grows up to 488~KB, which can again be reduced to 335~KB by removing unused classes from the standard Kotlin library.
Table~\ref{fig:memory_summary_static} summarizes the gain of the successive versions of KMF {\em w.r.t.} EMF.

\begin{table}
\vspace{-1em}
\center
{\renewcommand{\arraystretch}{1.2}
\begin{tabular}{|c| >{\centering\arraybackslash}p{1.7cm}| >{\centering\arraybackslash}p{2.2cm}| >{\centering\arraybackslash}p{2.5cm}|}
\hline Tool (Language) & Effective Code (KB) & Standalone Package (MB) & Reduced Standalone (MB)\\ 
\hline EMF (Java) & ~55 & 15 & No Data\\ 
\hline KMF (Kotlin) &  76 & 0,488 & 0,335 \\ 
\hline 
\end{tabular} 
}
\caption{EMF and KMF static memory footprint comparison}\label{fig:memory_summary_static}
\vspace{-1em}
\end{table}

Thanks to this drastically reduced footprint, KMF has successfully been used to generate APIs able to run on a large offer of JVM, including mobile and embedded ones: Dalvik~\footnote{http://www.dalvikvm.com/}, Avian~\footnote{http://oss.readytalk.com/avian/}, JamVM\footnote{http://jamvm.sourceforge.net/} or JavaSE for embedded Oracle Virtual Machine~\footnote{http://www.oracle.com/technetwork/java/embedded/downloads/javase/index.html}.

\paragraph{Dynamic memory footprint}
~\\
To further improve the dynamic memory usage, we paid more attention to the creation of temporary objects. One of the main consequence of that is the dropping of {\em scala.Option}~\cite{scala} (compared to our previous contribution).
The new version of KMF in Kotlin gets rid of these {\em scala.Option} because (1) these objects were among the most frequently created, generating a huge overhead; and (2) Kotlin does not support such a mechanism but proposes {\em Nullable} variables instead. This nullable mechanism relies on a static check at compile time introducing then no overhead at runtime.

Table~\ref{fig:memory_summary_dynamic} compares the sizes of heap memory used to create the experimental FSM model (presented in section~\ref{sec:sota} in EMF, KMF with Scala~\cite{Fouquet:2012fk} and finally, KMF with Kotlin.
\begin{table}[h!]
\center
{\renewcommand{\arraystretch}{1.2}
\begin{tabular}{|c| >{\centering\arraybackslash}p{1.8cm}| >{\centering\arraybackslash}p{2cm}| >{\centering\arraybackslash}p{2.2cm}|}
\hline Tool (Language) & EMF (Java) & KMF (Scala) & KMF (Kotlin)\\ 
\hline Heap Memory Used & 140~MB & 61~MB & 52,2~MB\\ 
\hline 
\end{tabular}
} 
\caption{EMF and KMF dynamic memory footprint comparison}\label{fig:memory_summary_dynamic}
\vspace{-1.5em}
\end{table}

This reduction of the memory used also reduces the need for garbage collection, because less objects are created and disposed.
Since the garbage collection in Java is a rather costly operation, this also participates to the efficiency of the overall approach.

\subsection{Save and Load}
\label{sec:contrib:saveLoad}

Load and save operations are two of the most used operations when dealing with models at runtime.
Indeed, any change in a model may have to be distributed to other devices taking part in the software system, or stored for history to be potentially restored later on.
The performance of these operations can thus generate a considerable overhead for little manipulations.

As described in section~\ref{sec:sota}, the reflexivity of loaders and the creation of temporary structures (objects) are two important bottlenecks.\\
~\\
{\bf Flat reflexivity}~\\
Thanks to the close world assumption, we decided in KMF to generate domain specific loaders and serializers. This flat generation removes all reflexive calls from the loading and saving processes making them more efficient.\\
{\bf Avoid temporary objects creation}~\\
Usually, (un)marshalling operations rely on an intermediate structure like a DOM to simplify the mapping from a structure in memory into a persistent structure ({\em e.g.} XML file). However, this implies the creation of complex temporary structures in memory.\\
To avoid this costly temporary structure, the loader works directly from the stream, and creates the final model elements directly. This is possible because each element of the stream is predictable (close world).
Similarly, KMF generates serializers that directly print in a stream buffer, to avoid unnecessary creation and concatenations of Strings.\\
{\bf Minimize class loadings}~\\
Previously~\cite{Fouquet:2012fk}, for each concept of the metamodel (domain), a specific class providing specific uniquely named loading/saving methods were generated and merged at compile time.
However, if this kind of generation makes a better organization of the code and eases the reading by a human being, it creates a lot of classes and requires, at runtime, many costly class-loading operations.
Thus, the new version of KMF presented in this paper groups all the loading (respectively saving) methods in a single file to drastically reduce the class loading time part of these operations.\\
{\bf Several exchange formats}~\\
The generative approach enables the generation of loaders and serializers for several standard formats.
At this time, KMF can generate I/O tools for the standard XMI and JSON, from files or streams (easier to load on some resource constrained platforms). They also provide a rich API to enable marshalling and unmarshalling in compressed formats.\\
~\\
{\bf Efficiency measurement}~\\
Table~\ref{fig:contrib::saveload:summary} summarizes the time required for loading and saving a model from/to a serialized formed with classical EMF/Java tools, former KMF version and the new KMF.
Again, the measures presented in the table are the results from the loading and serialization of the experimental FSM model presented in section~\ref{sec:sota}.

\begin{table}[h!]
\center
{
\renewcommand{\arraystretch}{1.2}
\begin{tabular}{|c| >{\centering\arraybackslash}p{2cm}| >{\centering\arraybackslash}p{2cm}|}
\hline Tool (Language) & Loading (ms) & Saving (ms)\\ 
\hline EMF (Java) & 1214 & 7021\\
\hline KMF (Scala) & 1193 & 799\\
\hline KMF (Kotlin) &  999 & 802\\
\hline 
\end{tabular} 
}
\caption{EMF and KMF load and save time}\label{fig:contrib::saveload:summary}
\vspace{-1.5em}
\end{table}


\subsection{Concurrent read/write access}
\label{sec:contrib:clone}
\vspace{-0.5em}
The protection against concurrent accesses can be fine-grained, on each individual model element, or coarse-grained, on the whole model.
This section first details fine-grained protection mechanisms for concurrent reads, then presents coarse-grained protection with efficient runtime clone operation for concurrent writes.\\

{\bf Concurrent shared model}~\\
To allow concurrent read access to model elements and relationships, the code generated by KMF (1) only uses standard Java thread-safe collections instead of ad-hoc EMF EList (or similar), which have issues on concurrent usages, and (2) exposes cloned immutable lists only via its public methods.\\

{\bf Model clone for independent modifications}~\\
To allow modifications of models, a simple strategy consists in providing each thread with a copy (clone) of the model, so that it can manipulate it independently from the other threads, with no need for a fine-grained synchronization.

Cloning is a very costly operation, basically because it duplicates entire structures in memory.
In the Java world, this means that every object from the model A yields a new equivalent object in the model A'.
To optimize the cloning of models, KMF relies on a flat reflexivity model to implement a very efficient, single pass, model-cloning operator. Moreover, all clones share the immutable Java objects like String resources to avoid duplication of invariant objects.\\

{\bf Towards immutable models}~\\
The KMF framework offers a method to turn individual model elements into read-only state.
The read-only state cannot be reverted to a read/write structure to ensure that no process will break this assumption while the model is being processed.
The only way to obtain a mutable structure is to duplicate the model into a new semantically identical model by cloning.\\
In addition, if an element is set as read-only, we also set all its contained elements and their references to read-only.
This ensures that no part of the sub-graph remains mutable when the root is immutable.
To do so, the generated API relies on the flat reflexivity and close world assumption to automatically call the $setReadOnly$ method on each sub element.\\

{\bf Partial clone operator}~\\
Cloning is a very costly operation, because it duplicates the graph structure in memory, similarly to the fork operation on a Linux process.
Despite the semantic of clone design implies this duplication, runtime optimizations can minimize the memory usage while keeping the right semantics. 
To this end, KMF proposes a partial clone operation.\\
The basic idea is to duplicate only the mutable parts of model, while sharing immutable parts between the original model and its clones.
Such a mechanism greatly improves the memory usage while leaving a lot of flexibility to the user. Indeed, mutable and immutable parts can be defined precisely using the \textit{readOnly} operation.

\begin{figure}[h!]
\center
\includegraphics[width=0.6\textwidth]{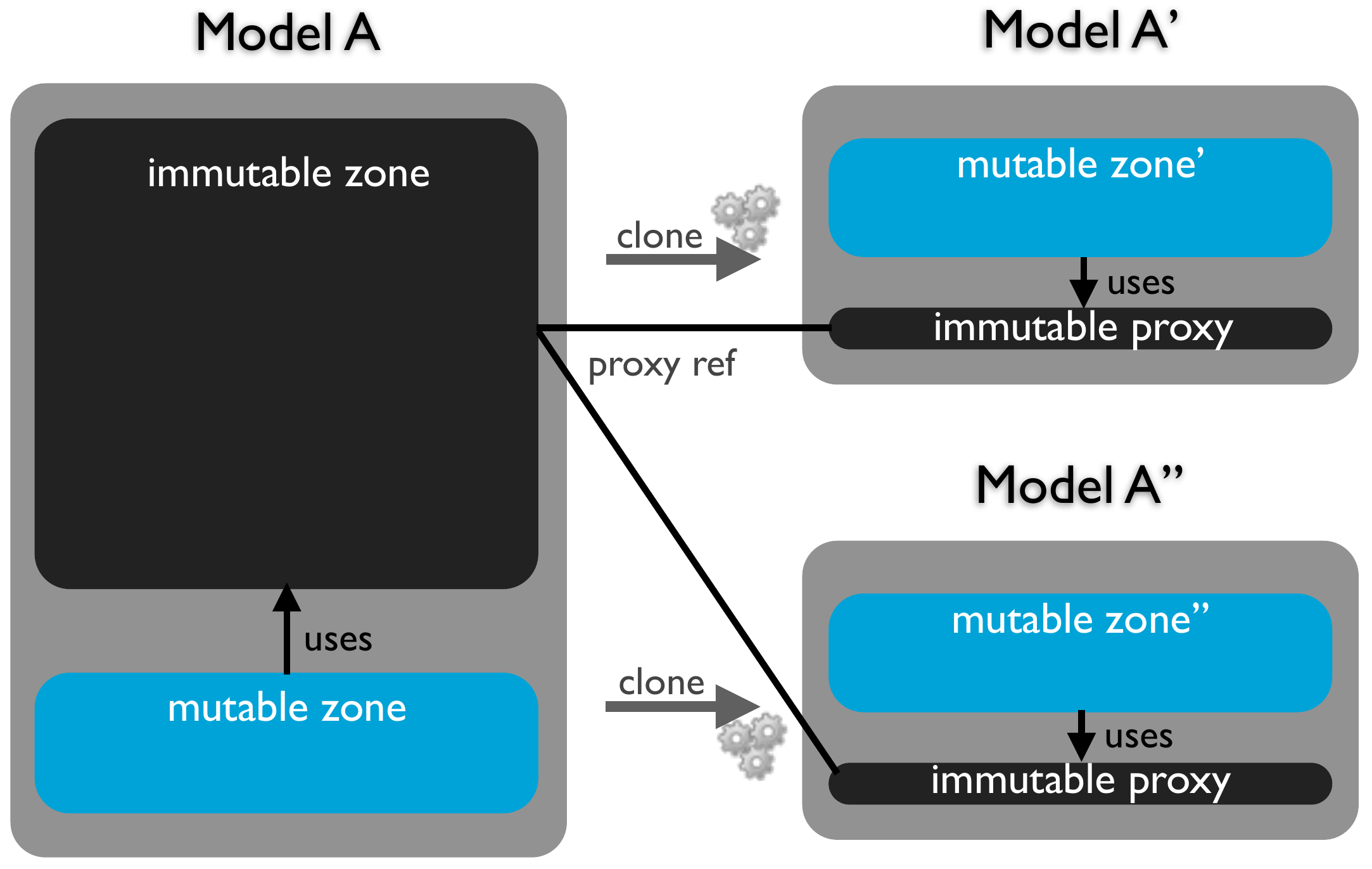} 
\caption{Partial clone operator illustration}\label{partialClone}
\vspace{-1em}
\end{figure}

Figure~\ref{partialClone} illustrates the partial clone operator where a model $A$ is cloned twice, and the immutable zone is shared between the clones $A'$ and $A''$.
The cloned model has a smaller size than the original model, due to the shared memory zone.
Moreover, since the clones reference the immutable zone, they remain consistent if the garbage collector frees the model $A$ ({\em i.e.:} the immutable zone will not be collected while at least one clone has a reference on it).

Finally the clone operation uses the $readOnly$ flag to prevent from navigating or cloning model elements that do not need to be duplicated, thus improving the cloning time.\\

{\bf Experimental validation}~\\
In order to evaluate this smart clone, the experiment takes a model $A$ from the Cloud domain, containing 400 nodes, and tries two strategies:
\begin{enumerate}
	\item One node is mutable and we perform a very specific modification, for example searching for component placement inside this node.
	\item all the 400 nodes are mutable, but we fix all static information (mainly related to provisioning).
\end{enumerate}

The hardware used in this experiment is a Macbook Pro (i7 processor 2.6Ghz and Oracle JVM 7) and results are presented in table~\ref{fig:partialCloneTab}.

\begin{table}[h!]
\center
{\renewcommand{\arraystretch}{1.2}
\begin{tabular}{|c| >{\centering\arraybackslash}p{2cm}| >{\centering\arraybackslash}p{2cm}|}
\hline experimental configuration  & time to clone (ms) & memory per model\\ 
\hline Full clone (1 mutable node) & 18 & 705kb\\ 
\hline Partial clone (1 mutable node) & 0.86 & $<$1kb\\ 
\hline Full clone (400 mutable nodes) & 19 & 705kb\\ 
\hline Partial clone (400 mutable nodes) & 5.36 & 304kb\\ 
\hline 
\end{tabular}
} 
\caption{Partial and Full clone comparison}\label{fig:partialCloneTab}
\vspace{-1em}
\end{table}

The clone time goes down from 18 ms to 0.86 ms, while the memory used for each clone drops from 705kb to less than 1kb per clone, which highlights the significant gain of memory.
Also, even if we take a larger zone of mutable elements, the gain is still significant with a time reduced from 19 ms to 5.36 ms and each clone takes twice as less memory as the original.
Improving clone performance enables the use of coarse grain strategies (model unit) instead of relying on costly synchronized model mutators.


\subsection{Efficient Model navigation : KMF Path}
\label{sec:contrib:path}

As illustrated by the example in section~\ref{sec:generic:nav}, activities to be carried on models@runtime need an efficient way to look up and navigate across model elements. 
In particular, during the model comparison steps, the merge and check operations require an efficient tool to reach a specific element in the model. 
To enable this efficient research, KMF leverages the notion of Unique Identifier from relational databases, which should be declared in the metamodel using the `id` field, already present in Ecore. 

This section introduces the Path Selector (KMFQL-PS) of the Kevoree Modeling Framework Query Language (KMFQL), which uses the id attribute specified in the metamodel, as the unique key to find a model element by following model relationships. 

\paragraph{Approach}~\\
Directly inspired by the select operator of relational databases and by XPath\cite{Consortium:2007fk}, KMFQL-PS defines a query syntax aligned with the MOF concepts. 
The navigation through a relationship can be achieved with the following expression: \textbf{relationName[IDAtrributeValue]}. 
This expression defines the \textbf{PATH} of an element in a MOF relationship.  
Several expressions can be chained to recursively navigate the nested models. 
Each expression is delimited by a \textbf{/}. 
It is thus possible to build a unique path to a model element, by chaining all sub-path expressions needed to navigate to it from the root model element, via the containment relationships. 

In our illustrating example, the ``name'' attribute is defined as the ID attribute of \textit{NamedElement}, and we know precisely where the component instance is hosted ({\em i.e.} the path to the model element in the model). 
The numerous nested loops presented in section~\ref{sec:sota} are now reduced to the piece of code presented in listing~\ref{lst:kmfql_ps}. 

\begin{scriptsize}
\begin{lstlisting}[frame=L, language=Java, caption={Collect of the FakeConsole component instance with KMFQL-PS.}, label={lst:kmfql_ps}]
ComponentInstance fConsole2 = model.findByPath("nodes[node6]"
	+ "/hosts[node7]/hosts[node8]/hosts[node4]"
	+ "/components[FakeConso380]",
	ComponentInstance.class);
\end{lstlisting}
\end{scriptsize}

\paragraph{Implementation details}~\\
KMF generates hash-tables in place of simple collections, for each relationship. 
Then, each time a model element is added in/removed from a relationship, it is stored in this hash-table. 
Hash keys are computed from the ID attributes, or are automatically generated if this ID is not defined. 

The keys in the hash-tables are equivalent to indexes in noSQL databases. Also, even if the use of hash-tables introduces a slight memory overhead, it considerably speeds up the resolution of the paths ({\em i.e.} the retrieval of a model element).
Moreover, the hash-tables allow to get rid of the high number of temporary objects created when looping, which favours the scalability on limited execution environments such as in the IoT domain.\\

{\bf Experimental validation\footnote{The code and more documentation about this experiment can be found within the KMF Github repository: https://github.com/dukeboard/kevoree-modeling-framework/blob/master/doc/kmf\_path.md}}~\\
We performed an evaluation of this feature on the model presented in Section~\ref{sec:generic:nav} (Figure~\ref{fig:kevoree_pathexample}). 
The APIs have been generated with EMF (2.7) and KMF, and the XMI model has been loaded prior. 
Then, as explained in section~\ref{sec:generic:nav}, we collected the ``FakeConsole380'' model element with both plain Java approach (listing~\ref{lst:plainJava}) and KMFQL-PS (listing~\ref{lst:kmfql_ps}).

For this resolution, the plain Java approach with \textbf{EMF} takes~\textbf{954$\mu$s} to execute, while the same resolution in \textbf{KMFQL-PS} takes~\textbf{37$\mu$s} on the same hardware (i7 processor 2.6Ghz and Oracle JVM 7). The resolution of already known model elements with KMFQL-PS is thus 25.7 times faster than the approach using nested for-loops applied on EMF models.
Also the algorithmic complexity has been significantly reduces compared to the nested for-loops approach (see section~\ref{sec:generic:nav}), to O($\log(n) \times \log(o) \times \log(p) \times \log(q) \times \log(r)$), which confers a better scalability to the approach.

\vspace{-1em}
\section{Evaluations}
\label{sec:evaluation}

\subsection{Overview}
Each requirement has previously been evaluated in the dedicated sub-sections of Section~\ref{sec:contrib}. 
The aim of this section is to provide more insights on the performances of KMF over different models, from different domains, with different sizes and features.

\subsection{Experimental protocol}
The experimental protocol is composed by the following steps, which compose a standard process when using models at runtime. For each model / domain:
\begin{enumerate}
	\item The model, previously serialized in the XMI standard format, is loaded from a file (repeated 10 time),
	\item The model is then cloned (100 repetitions) to prepare a search/comparison,
	\item One clone and the original model are compared (1 time), element by element,
	\item Finally, the clone is serialized (10 repetitions) in a file, in XMI format.
\end{enumerate}

This process is repeated on models of various sizes, conforming to three representative metamodels:

\begin{enumerate}
	\item {\bf Kevoree}: the platform leveraging Model@Runtime previously detailed in this paper~\cite{fouquet:hal-00688707,Fouquet:2012:DCM:2304736.2304759}
	\item {\bf Kermeta}: a model-oriented language developed by the Triskell Team to define the operational semantics of models~\cite{kermeta}
	\item {\bf ThingML}: an operational DSL dedicated to the Internet of Thing (embedded systems connected to the Internet)~\cite{thingml}
\end{enumerate}

All the experiments are run on the same hardware as the one used to evaluate EMF in the requirements section ({\em i.e.}  MacBook pro, Intel Core i7 2.6Go, 16Go RAM and SSD drive).
The JVM used is the Oracle standard JVM 7 and all files and code used during the experiment are publicly available on the project Github\footnote{https://github.com/dukeboard/kevoree-modeling-framework/tree/master/sosym-evaluations}.

\subsection{Results}

\begin{figure}
\center
\includegraphics[width=1\textwidth]{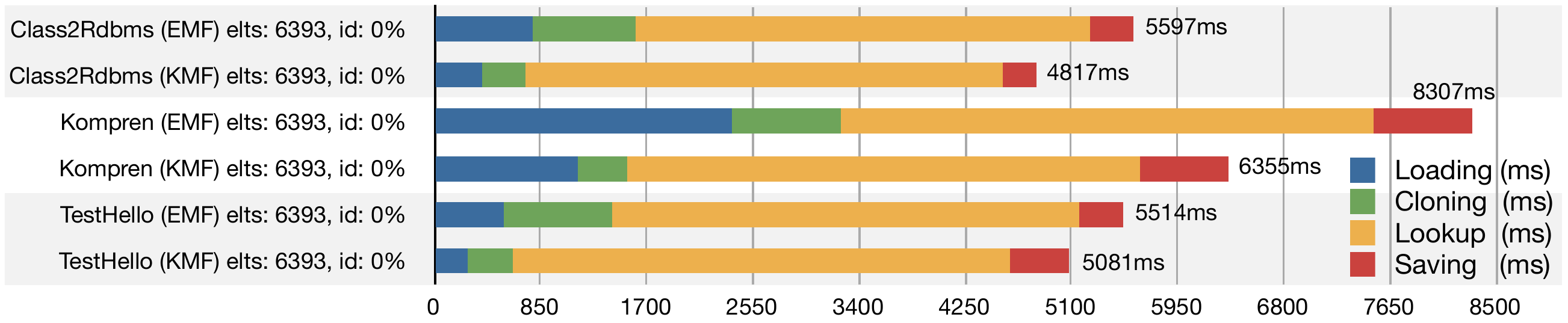} 
\caption{Evaluation of KMF and EMF on Kermeta models}\label{eval:KermetaComparison:fig}
\vspace{-0.5em}
\end{figure}

\begin{figure}
\center
\includegraphics[width=1\textwidth]{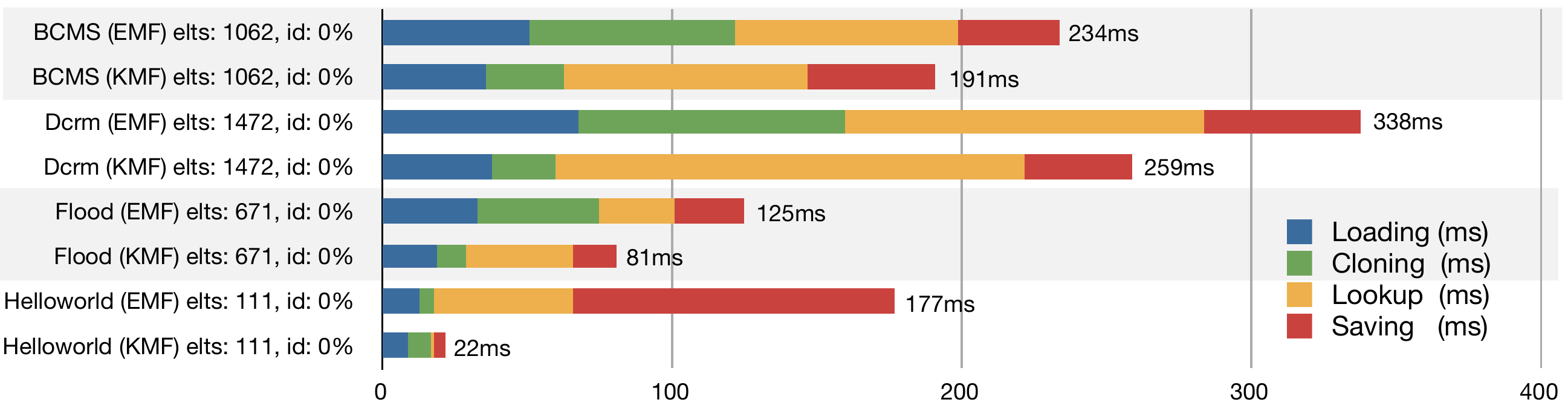} 
\caption{Evaluation KMF and EMF on ThingML models}\label{eval:ThingMLComparison:fig}
\vspace{-1em}
\end{figure}

{\bf Generalization of performance improvement on different metamodels}~\\
The benchmark is executed on a first set of models conforming to Kermeta and ThingML metamodels.
Their sizes span from 100 to 6000 elements, and have different structures implied by their metamodels.\\
The results of KMF and EMF when dealing with Kermeta and ThingML models are graphed in Figure~\ref{eval:KermetaComparison:fig} and Figure~\ref{eval:ThingMLComparison:fig}.\\
It is important to note that the metamodels (as well as the models) of Kermeta and ThingML are used as-is and have not been modified to leverage the specific capabilities of KMF (in particular the definition of IDs).

Overall, the code generated by KMF tends to be more efficient than the one with EMF (in addition to other benefits such as thread safety, discussed in Section~\ref{sec:contrib}), with an average gain of 28\% on our benchmarks. 
This confirms that KMF is generalizable, even if the metamodels have not been optimized for KMF generation. \\
Also, we observe in these experiments that the optimization of the generated API can not go beyond the limit of 30\% of gain, because of the graph complexity of models.
This advocates for the need of new features to reduce more significantly the algorithmic complexity, such as partial clone and query language which are evaluated in the next two sections.\\

{\bf Performance evaluation of the KMFQL-PS}~\\
The metamodels of Kermeta and ThingML do not define IDs and thus, do not leverage KMF optimized paths.
A second set of models, describing the topology of cloud stacks (software, platform and infrastructure), is created.
These models have different size, ranging from 876 to 33644 elements.
This second experiment varies the use of ID definitions and the size of the models in order to evaluate the impact of the new KFM concepts.

\begin{figure}[h!]
\vspace{-1em}
\center
\includegraphics[width=1\textwidth]{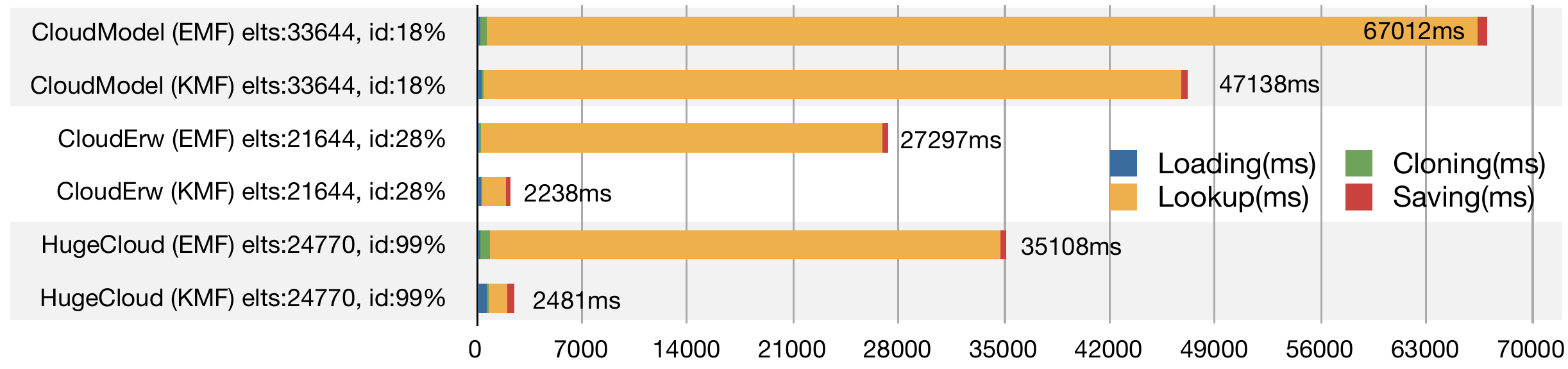} 
\caption{KMF results on Kevoree models (models@runtime)}\label{eval:SecondPack}
\vspace{-1.5em}
\end{figure}

Three cloud models have been selected depending of their percentage of ID definition (18, 28 and 99\%).
As a reminder the KMF lookup method reuse ID information to speed-up the resolution.
The results are presented in Figure~\ref{eval:SecondPack}.

This evaluation clearly highlights the benefit of defining ID attributes on model elements.
The code generated with KMF already allows gaining time compared to the EMF approach, with a very low percentage of ID definition (18\%) as illustrated by the first grey line in figure~\ref{eval:SecondPack}.
This gain is made even more important as the percentage of model elements that have an ID grows (grey lines 2 and 3 of the figure, with respectively 28\% and 99\% of elements with IDs).
The most significant gain is clearly on the lookup resolution, extensively used for comparison, merging, composition, {\em etc}.
From a wider point of view, the global process of loading, cloning, searching and saving is reduced from 35s, with EMF, to less than 2.5s with the optimized KMF.\\

\begin{figure}[hb]
\vspace{-1em}
\centering
\begin{minipage}{.45\textwidth}
  \centering
  \includegraphics[width=0.90\linewidth]{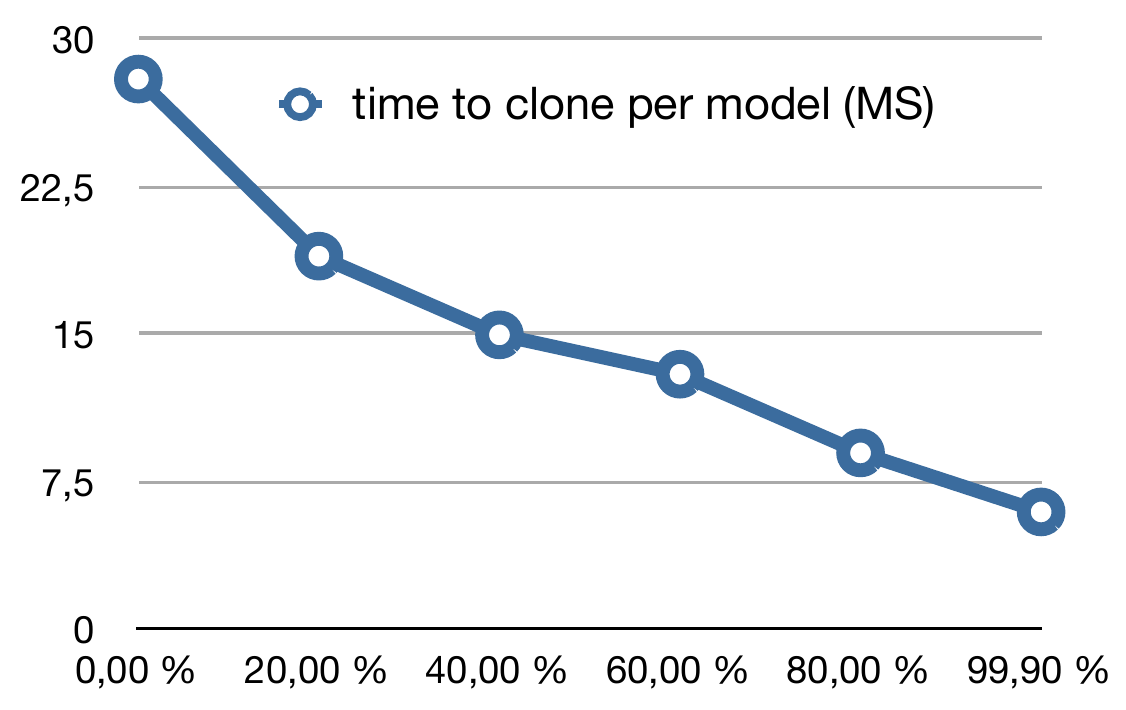}
  \captionof{figure}{Time to clone w.r.t. percentage of read-only elements in the model}
  \label{fig:readOnlyComparison:fig}
\end{minipage}%
\hspace{1em}
\begin{minipage}{.45\textwidth}
  \centering
  \includegraphics[width=0.99\linewidth]{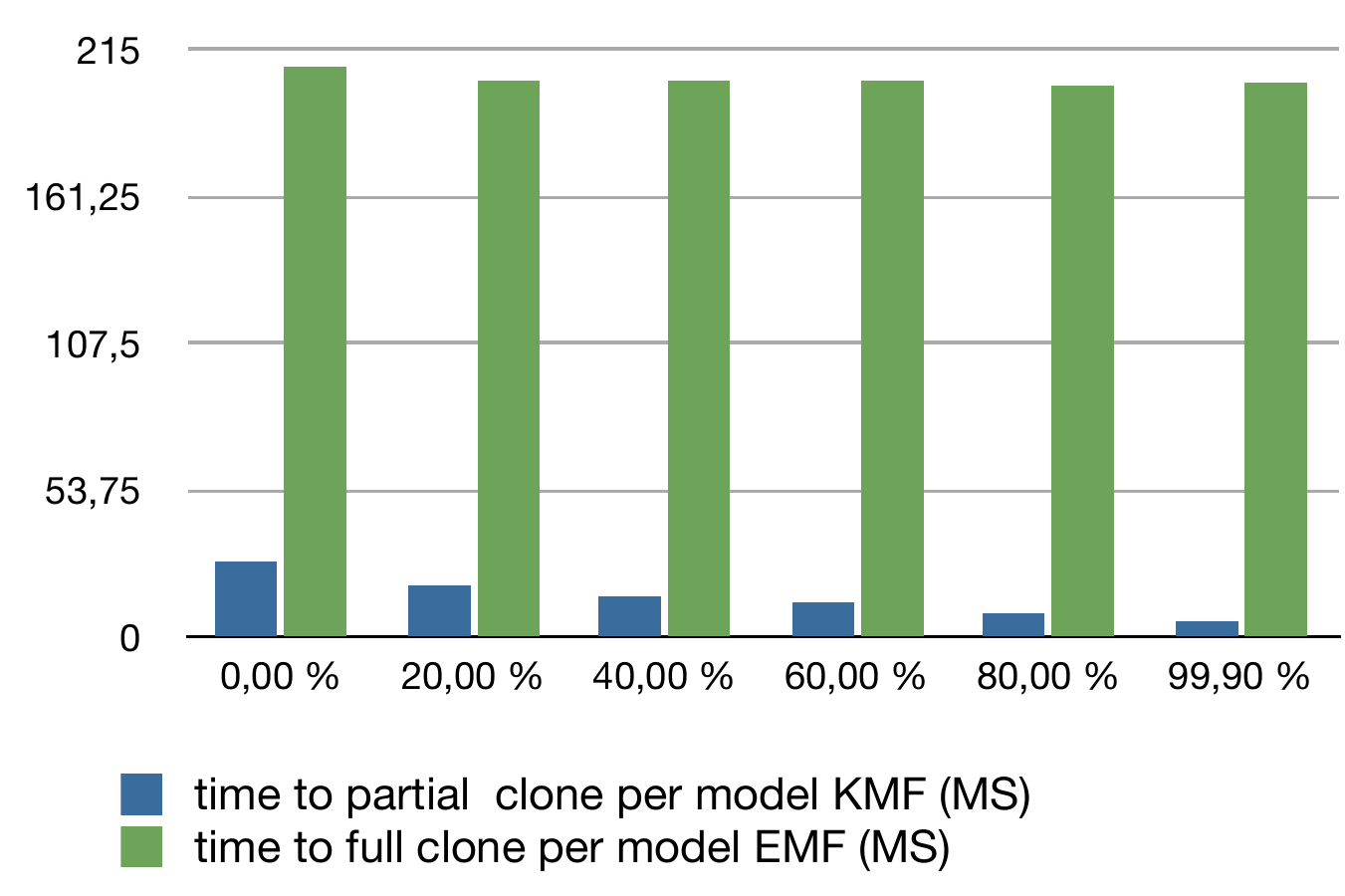}
  \captionof{figure}{Time to clone w.r.t. percentage of read-only elements in the model}
  \label{fig:readOnlyComparison:fig2}
\end{minipage}
\end{figure}

{\bf Performance evaluation of the Partial Cloning mechanism}~\\
The final experimentation aims at assessing the improvements brought by the use of partial clones.
To highlight the gain, a cloud model is used.
In this model, several nodes are containing other nodes and components.
The \textit{readOnly} option offered by KMF is progressively applied on 20, 40, 60, 80 and 99\% of the model.
For each percentage, the clone in KMF is compared with the clone in EMF.\\ 
The results are graphed in figures~\ref{fig:readOnlyComparison:fig} and \ref{fig:readOnlyComparison:fig2}.
We observe that the reduction of time to clone the model is close to be inversely proportional to the percentage of readOnly elements({\em i.e.} the more readOnly elements, the less time required).\\
The Figure~\ref{fig:readOnlyComparison:fig2} presents the result of the same experiment with EMF without any readOnly optimization.
Such results are taken only for reference, because EMF can only perform a full clone, time is then not impacted by the percentage of readOnly elements, and remains to a value around 204ms.

The results clearly show that the individual features of KMF presented and preliminary assessed in Section{sec:contrib}, also significantly improve the performances of a complete models@runtime process. The performance gains compared to the first experiment clearly emphasize the role of the KMF-specific features (partial cloning and path) in this performance improvement.

\section{Conclusion}
\label{sec:conclusion}
This paper discusses the needs for adapting the way of using MDE techniques created for design time activities, to support the use of models at runtime. 
To this end, several requirements are described, and the {\em de facto} standard in the MDE community, {\em i.e.} the Eclipse Modeling Framework (EMF), is evaluated against these requirements. 
After this evaluation, this paper presents the Kevoree Modeling Framework (KMF) as an alternative solution to create modeling tools more suitable for runtime purposes. 

This paper has presented the evolutions of the KMF generated code, according to what was presented in \cite{Fouquet:2012fk}.
The new version of KMF offers better performances with a reduced footprint, allowing to embed KMF more easily in resource-constrained JVM such as Android or Embedded Java. 

This new version of KMF also introduces several new features. 
KMF now provides a way to handle concurrency at the level of the model, with the ability to define  fragments as read-only. 
This latter option is particularly useful when combined with the new clone operator, making it possible to factorize read-only fragments among model clones (hence saving memory) while the mutable part is specific to each clone.
Finally, KMF also provides a query language to directly access a model element by providing a path, inspired by well-established work in the database domain.  

This new version of KMF has been compared to EMF and the initial version of KMF~\cite{Fouquet:2012fk}.
These evaluations clearly show the gain offered by the KMF generated API in several domains, with an average gain of 28\% in time with respect to the EMF implementation.
Then we highlighted the need to rethink some modeling operators such as cloning and lookup to make them more efficient at runtime.
The empirical evaluation clearly shows that the use of unique identifiers and KMFQL-PS saves lots of computation time.

This work finds several application cases in domains like Internet of Things, Cloud management or dynamic adaptation of software systems. 
The opportunities of application are also augmented by the good performances of the KMF generated code and the reduction of the memory required compared to EMF generated code.
These good properties should enable a wider use of models at runtime in industrial scale projects.

In future work, we plan to port KMF to JavaScript, to make the generated API available directly in web-browsers, making it possible to develop web-based, massively distributed, collaborative modeling environment.
Finally, also plan use and extend KMF in the context of Big Data models.
Because of their particularly big size, these models cannot be entirely loaded in memory and require seamless memory swapping between memory and disk.
Based on existing works~\cite{pagan2011morsa,Barmpis:2012:CAD:2467307.2467314}, several approaches are currently being tested to offer MDE WITH seamless manipulation solutions for these big models, mostly inspired by noSQL database approaches.


%
%

\bibliographystyle{plain}

\begin{footnotesize}

\bibliography{biblio}
\end{footnotesize}

\end{document}